\documentclass[12pt]{article}

\usepackage{latexsym}
\usepackage{multirow}
\usepackage[linesnumbered,ruled,algonl,vlined]{algorithm2e}

\usepackage{amsfonts}

\usepackage{times}
\usepackage{amsmath}

\usepackage{balance}  
\usepackage{graphics} 
\usepackage{times}    
\usepackage{url}      
\usepackage{multirow}
\usepackage{subfigure}
\usepackage{paralist}

\usepackage{graphicx}
\def\squareforqed{\hbox{${\Box}$}}
\def\qed{\ifmmode\squareforqed\else{\unskip\nobreak\hfil
\penalty50\hskip1em\null\nobreak\hfil\squareforqed
\parfillskip=0pt\finalhyphendemerits=0\endgraf}\fi
\vskip 6pt}

\newcounter{linenumber}
        {\end{list}}

\newcounter{tmp}

\newcommand\blankpagestyle{empty}

\def\cleardoublepage{\clearpage
  \if@twoside \ifodd\c@page
  \else
    \hbox{}\thispagestyle{\blankpagestyle}\newpage
    \if@twocolumn\hbox{}\newpage\fi
  \fi\fi
}

\newcommand{\eat}[1]{}

\date{}

\title{Impact of Imbalance Usage of Social Networking Sites on Families}

\author{%
{Anika Anwar{\small $^{1}$} \hspace{0.3cm}Ishrat Ahmed{\small $^{2}$} \hspace{0.3cm}Tanzima Hashem{\small $^{3}$} \hspace{0.3cm} Jalal Mahmud{\small $^{4}$}}%
\vspace{1.6mm}\\
\fontsize{10}{10}
$~^{1,2,3}$Dept. of CSE, Bangladesh University of Engineering and Technology, Bangladesh,\\
$~^4$IBM Research-Almaden, San Jose, CA, USA\\
\fontsize{9}{9}\selectfont\ttfamily\upshape
\{$^1$anika.anwar,$^2$ishratahmedmren\}@yahoo.com, $^3$tanzimahashem@cse.buet.ac.bd, $^4$jumahmud@us.ibm.com\\}

\begin{document}

\maketitle
\begin{abstract}
With the proliferation of social networking sites (SNSs) such as Facebook and Google+, investigating the impact of SNSs on our lives has become an important research area in recent years. Though SNS usage plays a key role in connecting people with friends and families from distant places, SNSs also bring concern for families. We focus on imbalance SNS usage, i.e., an individual remains busy in using SNSs when her family member is expecting to spend time with her. More specifically, we investigate the cause and pattern of imbalance SNS usage and how the emotion of family members may become affected, if they use SNSs in an imbalanced way in a regular manner. This paper is the first attempt to identify the relationship between an individual's imbalance SNS usage and the emotion of her family member in the context of a developing country.
\end{abstract}

\section{Introduction}
Social networking sites (SNSs) such as Facebook, Google+, and Twitter, have enabled users to remain virtually connected with their families and friends at any time and from any places around the world. Large varieties of social-networking applications are continuously attracting a large number of people to use SNSs. Along with the benefits, SNSs sometimes also bring concerns to our family lives. Family members, who live in the same house, expect to spend some time together in gossiping, watching television and outing in holidays. Researchers have also demonstrated the importance of family leisure for family cohesion \cite{orthner}. However, nowadays, it is quite common, that people remain busy in SNSs when their family members are expecting to spend time with them, which we call \emph{imbalance SNS usage}. This research is the first attempt to investigate how an individual's imbalance SNS usage affects psychological wellbeing of her family members.

Family members, especially adolescents~\cite{Jennifer}, often use SNSs excessively. In our initial study, we observe that an individual becomes really frustrated if her spouse remains always busy in social networking instead of passing time with her. Imbalance SNS usage becomes a more significant concern, when family members use SNSs in an imbalanced way \emph{in a regular manner}. The most common form of imbalance SNS usage happens when one family member is active in SNSs whereas the other is not. Imbalance SNS usage not only reduces family time, it may also prevent an individual, who does not have access to SNSs, from participating in different family discussion and activities. The imbalance SNS usage can also arise even if both family members have SNS accounts; both members are using SNSs at different times when the other member is expecting to pass time with her.

It is true that imbalance may also arise on families from other activities. For example, one may remain busy in gardening or any household chore (e.g., cooking) or any other digital activity while her family member is expecting to pass time with her. However, gardening or any other hobby or digital activity has not become as ubiquitous as social networking. On the other hand, we argue that a household chore is essential for our living and the time spent in household activities is limited. Thus, the problem of imbalance SNS usage requires special attention from researchers to make SNSs more enjoyable for users.

To know the attitude and experience of individuals due to the imbalance SNS usage of other members in families, we began with interviewing people, who suffer from imbalance SNS usage of their family members. From interviews, we identified that annoyance, sadness and loneliness are the most commonly observed emotions of people due to the imbalance SNS usage of their family members. Based on the feedback from interviews and statistical analysis, we prepared a group of correlated questionnaire to measure different emotion levels and ran an anonymized survey among 53 families. We observe that imbalance occurs between two family members in using SNSs mostly (71\% cases) because one of them is active in SNSs but the other is not and for approximately 80\% cases, young SNS users are responsible for causing imbalance with their family members.

Our findings validates the fact that the negative impact of imbalance usage of SNSs is significant on family lives and necessitates to make people aware of this fact so that people can participate in SNSs with their family members in a balanced way.

\subsection{Related Work}
Recent researches have focused on different aspects of SNSs that relate with families~\cite{BurkeAM13,Social,kusss}. In~\cite{Stein,Jean,Robert,Charles}, researchers have investigated how Internet and SNSs psychologically affect human being. The work in~\cite{nettime} has focused on how family members negotiates in sharing resources to access the net. Our work is different because we investigate how one's imbalance SNS usage can affect the psychological wellbeing of other family members.

\subsection{Research Questions}
Our research is guided by a few research questions. We aim to investigate the impact of imbalance SNS usage on families, which leads to the following research question:

\emph{\\\textbf{RQ1} How does imbalance SNS usage occur in families? How does the imbalance SNS usage of a user affect psychological wellbeing of her family members?}
\\

In a study~\cite{DBLP:conf/ictd/RedaSTLN12}, researcher have shown that young generation and males use SNSs more than the elderly people and females in developing countries, respectively. Therefore we investigate:

\emph{\\\textbf{RQ2} Do Male and young generation have a vital role in causing imbalance SNS usage?}
\\

Finally, we consider the relationship between imbalance SNS usage and demographic variables:

\emph{\\\textbf{RQ3} Do the impact of imbalance SNS usage vary with demographic variables such as age, gender and relationship?}

\section{Method}

To investigate the cause and pattern of imbalance SNS usage and its effect on family members, we conducted a study in a metropolitan city of a developing country, Bangladesh. We chose a developing country because the probability of imbalance SNS usage is higher in developing countries due to existing digital divide in using technology~\cite{DBLP:conf/ictd/RedaSTLN12}.

Since our goal is to know the impact of imbalance SNS usage, in our data collection process, we had to first identify sample population whose family members use SNSs in an imbalanced way. To prepare a questionnaire for survey, we first interviewed 15 people, who were suffering from the imbalance SNS usage of their family members, to know how imbalance occurs and how they feel when one of her family member remains busy in SNSs instead of passing time with her. Based on these interviews, we prepared a questionnaire and conducted a survey among 53 pairs of family members to know more about their daily life activities and emotions that become affected due to the imbalance SNS usage.

\subsection{Interviews}
Interview participants were recruited using word of mouth approach. The interviewer asked a set of general questions related to SNS usage to a selected set of friends from their known circles and identify 15 interview participants whose family members' SNS usage pattern match the concept of imbalance SNS usage, i.e., the participants' family members use SNSs when participants expect to spend time with them.

The interviewers observed that when they asked the participants about how they feel when one of her family member remains busy in SNSs instead of passing time with her, they hesitated to share their experience with their intimate family members to the interviewers. To make the participant comfortable, the interviewer ensured the participants that their interviews would remain anonymous and would be used for research purpose. The interviewer took the interview of each participant separately and most of the cases at the participant's home. The interviewer took hand notes, which were also cross checked by the participants at the end of interviews. Sample interviews and the demographic information of participants are included in~\ref{sample_interview}.

From interviews, we found that the most common negative emotion that people suffer from is annoyance. The participants also seemed to be sad, and sometimes lonely when their family members use SNSs in an imbalanced way. We also asked them whether they think that they are depressed due to the imbalance usage of their family members and the answers were ``No" for most of the participants with a very few exception. We observed that the level of annoyance, sadness and loneliness varied among the participants for different levels of imbalance SNS usage of their family members.

\subsection{Surveys}
In our interviews, we found a large number of people who suffered from the imbalance SNS usage of their family members but hesitated to have a face to face discussion about their sufferings. Therefore, for our analysis, we conduct an anonymous paper survey to know the cause and effect of imbalance SNS usage from more participants. We did not go for an online survey because we observed that people who suffer from imbalance SNS usage often do not have access to the Internet due to lack of knowledge in technology.

Based on interviews, we prepared a questionnaire to identify whether people are suffering from commonly affected emotions, i.e., annoyance, sadness and loneliness due to the imbalance SNS usage of their family members. Our questionnaire also asked for demographic information of the participants such as age, occupation, gender, and relationship with their partner in the survey. The eligibility to participate in our paper survey were (i) family members have to participate in pair, who live in the same house (ii) at least one participant in the pair uses SNSs. Each participant in the pair answered questionnaire independently. We conducted the survey in pair to ensure reliability, we asked several same questions to both participants and checked whether their answers matched.

Since a recent study shows that young generation use SNSs excessively~\cite{DBLP:conf/ictd/RedaSTLN12} in developing countries, we predict that the probability of using SNSs in an imbalanced way is high for young generation. Therefore, we distributed the questionnaire to students of four universities. In addition, we also distributed questionnaire in our known circle. 53 pairs returned the filled survey.

\begin{small}
\begin{table}[htbp]
\centering
\caption{Computation of the Level of Annoyance}
\begin{small}
\begin{tabular}{|p {9cm}||p {1.5cm}|c|} \hline
I feel lonely because & Replies & Values\\ \hline
My survey partner uses SNSs when I am expecting to spend time with my survey partner. & Agree & 1  \\
\hline
Using SNSs, my survey partner shows stuffs (e.g. photos) to other family members and can see shared stuffs (e.g. photos) from other family members. I cannot see those stuffs (e.g., photos). & Neutral & 2\\ \hline
\hline
My survey partner uses SNSs for a long period of time continuously. & Agree & 1\\ \hline
My survey partner logs on to social networking sites instead of spending time with me when s/he has nothing to do. & Disagree & -1\\ \hline
My survey partner uses SNSs in parallel when s/he is spending time with me, e.g. gossiping with me and using social networking sites at the same time. & Strongly Agree & 0  \\ \hline
\hline
\multicolumn{2}{|l|}{Level of Loneliness } &  3\\ \hline
\end{tabular}
\end{small}
\label{tab:measuredetails}
\vspace{-3mm}
\end{table}
\end{small}

\subsubsection{Measures}
We could not use standard questionnaire to measure annoyance, sadness or loneliness because they are developed to measure emotions caused by any reason in general, whereas our goal is to determine whether annoyance, sadness, and loneliness are caused due to the imbalance SNS usage by one of the family member. To avoid biasedness, we did not directly ask the participants about whether they feel annoyed, sad or lonely due to the imbalance SNS usage. Specifically, in every statement related to these emotions in our questionnaire, we explicitly mention that we are referring to the time of imbalance SNS usage, i.e., a family member is using SNSs when the survey participant is expecting to spend time with her. We provided a group of correlated questions for each emotion.

The answer of each question related to annoyance, sadness, and loneliness were computed based on answers given on 5-pt Likert scale ranging from strongly agree to strongly disagree. We associated numerical values 2, 1, 0, -1, -2 with strongly  agree, agree, neutral, disagree, strongly disagree, respectively. Without loss of generality, Table~\ref{tab:measuredetails} shows an example to compute the level of loneliness.  We normalized the range of each emotion level between 0 to 1. Any emotion level having a value greater than 0.5 represents that the individual is suffering from the emotion.

The participants are allowed to keep the answer blank or select neutral if they feel that the fact stated in a statement is not true (applicable) for them. For example, it may happen that the event mentioned in the last statement in Table~\ref{tab:measuredetails} is not true for a participant and thus, there is no chance to become annoyed for this reason. On the other hand, sometimes it may happen that an event stated in a statement is true for a participant but it does not affect the emotional state of the participant. In such a case, the participant can opt for disagree or strongly disagree.

We computed Cronbach's Alpha ($\alpha$), which is commonly used to assess the reliability of questionnaire when multiple Likert questions form a scale. The values of $\alpha$ for annoyance, sadness and loneliness are 0.7521, 0.7008, and 0.7725, respectively, which show good reliability of our questions.
\begin{small}
\begin{table}[htbp]
\vspace{-1mm}
\begin{center}
\begin{small}
\caption{Demographic Data}
\begin{tabular}{|l|l|l|l|l|l|}
\hline
\multicolumn{2}{|l}{Age } & \multicolumn{2}{|l}{Gender }                  & \multicolumn{2}{|l|}{Relationship} \\ \hline
16-25         & 46         & Male                  & 38                  &  Parent-Child           & 28         \\ \hline
26 -35          & 11          & \multirow{2}{*}{Female} & \multirow{2}{*}{52} & Sibling          & 14       \\ \cline{1-2} \cline{5-6}
36-45           & 7        &                    &                    & Spouse           & 3         \\ \hline
46-55          & 20          & \multicolumn{4}{|l|}{In 71 \% cases one participant uses}                                          \\ \cline{1-2}
56-65          & 6        & \multicolumn{4}{|l|}{SNS and other does not}                                            \\ \cline{1-2} \hline
\end{tabular}
\label{tab:demo}
\vspace{-6mm}
\end{small}
\end{center}
\end{table}
\end{small}

\subsubsection{Survey Participant Demographic}

From 53 pairs, i.e., 106 participants in our survey, we discarded 8 pairs due to data mismatch and missing information and finally, we had 45 pairs.  Participants were from diverse background and occupations. Table~\ref{tab:demo} shows the demographic data of participants. The ethnic distribution of participants ensures that the research is not confined to a particular group. From our SNS usage related questions, we identify that among 90 participants (i.e., 45 pairs), survey partners of 62 participants use SNSs in an imbalanced way.

\begin{small}
\begin{table}[htbp]
\centering
\begin{small}
\caption{Reasons for not using SNSs}
\begin{tabular}{|c||c|} \hline
Reasons & Participant Percentage\\ \hline
Don't know how to use it & 36.84\% \\ \hline
Don't have time to use it & 28.10\% \\ \hline
Don't want or like it & 26.31\%\\ \hline
Don't have the resources & 7.01\%\\ \hline
Afraid of privacy violations & 1.75\% \\
\hline\end{tabular}
\label{tab:reasons1}
\end{small}
\end{table}
\end{small}

\subsubsection{Result}
In our survey, we observed that in 71\% of the cases imbalance occurs because one member is active in SNSs and the other is not. The participants who do not use SNSs are mostly (64\%) female. Table~\ref{tab:reasons1} shows the reasons of not using SNSs.

\begin{figure}[htbp]
    \centering
    \begin{tabular}{cc}
      \hspace{-12mm}
      \resizebox{50mm}{!}{\includegraphics{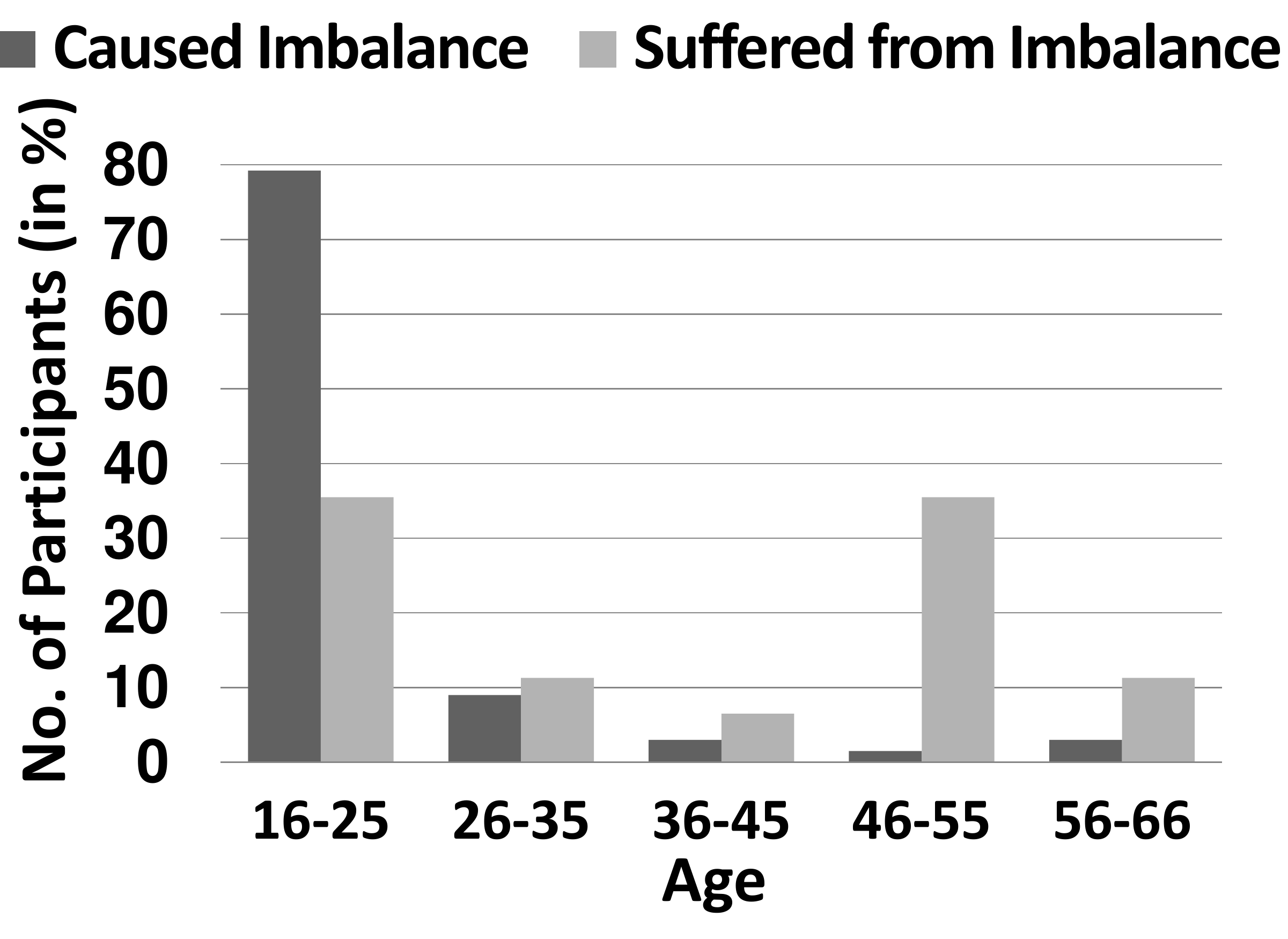}} &
      \hspace{-0mm}
      \resizebox{50mm}{!}{\includegraphics{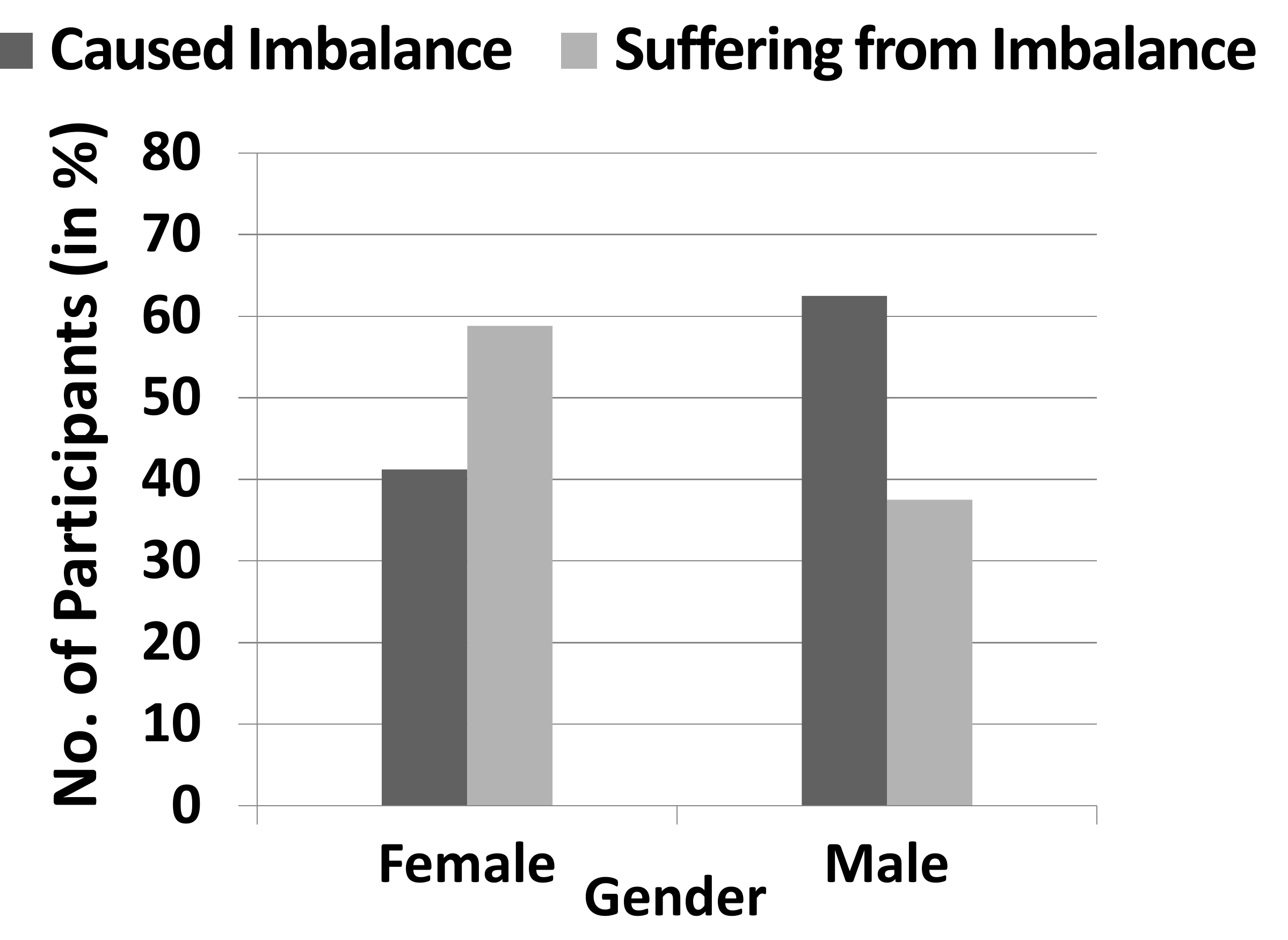}} \\
      \scriptsize{(a)}  & \hspace{-0mm}\scriptsize{(b)} \\
        \end{tabular}
    \caption{No. of participants (\%), who cause and who suffer from imbalance SNS usage}
    \label{fig:UA}
\end{figure}

In Figure~\ref{fig:UA}(a), we observe that young generation, who are in the age range 16--25 (79.24\%), are mainly responsible for imbalance SNS usage. People from age range 16--25 and 46--55 (35.48\%) are more affected from imbalance SNS usage of their family members. Figure~\ref{fig:UA}(b) shows that male are causing more imbalance whereas female are suffering more from the imbalance SNS usage.

From our survey, we find that the percentage of participants, who are annoyed, lonely and sad are 62.29\%, 42.62\%, and 40.9\%, respectively, and 87.1\% users are suffering from at least one negative emotion. This results validate that imbalance SNS usage has significant impact on families.

\begin{figure}[htbp]
    \centering
    \hspace{-5mm}
     \resizebox{80mm}{!}{\includegraphics{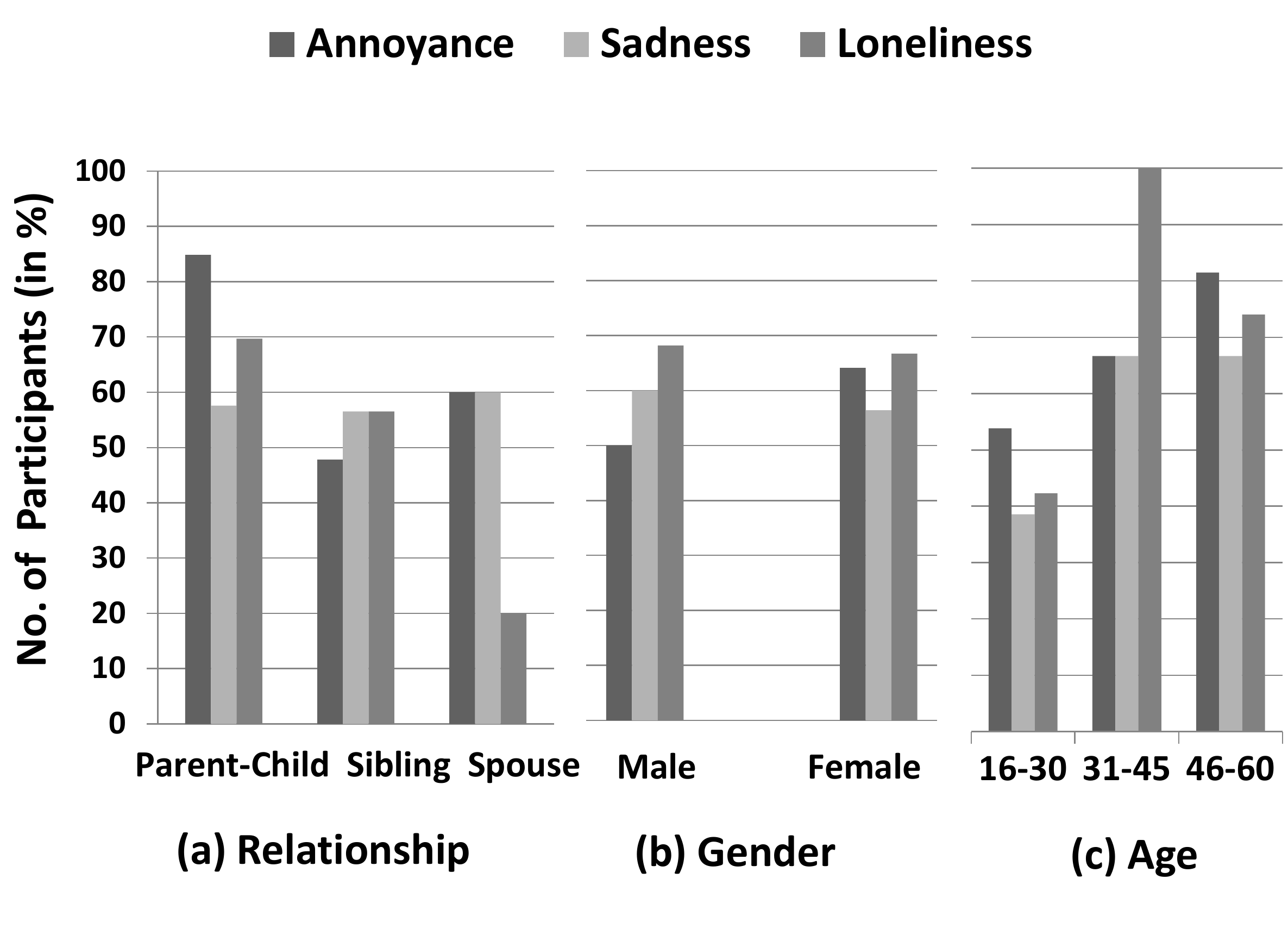}}
    \caption{No. of Participants (in \%)  based on (a) Gender, (b) Relationship and (c) Age}
    \label{fig:ratio}
\end{figure}

Demographic variables have role over the relationship between the imbalanced SNS usage and emotions. Figure~\ref{fig:ratio} shows the percentage of participants (\%), who suffer from different emotions based on relationship type, gender and age. We observe that gender does not have much impact on emotion levels, parents are more annoyed and all people in middle age group, who suffer from imbalance SNS usage, are lonely.

\section{Discussion}

We analyzed the anonymized data collected in the survey and attempted to find the answers of our research questions from the result of our analysis in the context of a developing country.

For our first research question, RQ1, we find that the main cause of imbalance SNS usage between two family members is one is active in SNSs and other is not. From our survey data we see that the main reason of not using SNSs is the participants do not know how to use it. Interview participants also shared their experiences about the pattern of imbalance SNS usage in their lives. Participants who are victim of imbalance SNS usage of their family members reported that their family members spend significant amount of time in SNSs when they are back from work. Some participants mentioned that their family members often use SNSs while they are in a family gathering. Some parents, especially stay-at-home mothers are alone whole day and expect to spend some time when her child is back from school. They would like to have a nice hangout or meal together, but in reality, children remain busy in SNSs. Both interview and survey data confirm that imbalance SNS usage has psychological affect on family members and commonly observed negative emotions among family members are loneliness, sadness and annoyance. Though a family member may become lonely/sad/annoyed also from others activities of her partner, in both interview and survey we explicitly mentioned that we are referring to the moment when their parters use SNSs instead of passing time with them.

In response to RQ2, our results show that female and people from age group 46--55 are most affected from imbalance SNS usage and young people are mainly using SNSs in an imbalanced way. On the other hand, for RQ3, we find that age and relationship have affect on the relationship between imbalance SNS usage and the affected psychological variables.

\section{Limitation and Future Work}
This paper presents an analysis on the cause and effect of imbalance SNS usage through a survey. The survey participants are from different educational and professional background. However, the survey still lacks some diversity. The survey was conducted in a metropolitan city of a developing country. In the future, we will conduct our survey in rural areas and in different countries.

We did an anonymous survey, thus we have no option to track them and repeat the survey to know the level of imbalance SNS usage and emotional state after a specific period of time, which would allow us to know the change of emotion due to the imbalance SNS usage over time. In future, we plan to perform a periodic study to examine how emotions of people change over time with the change of the level of imbalance SNS usage.

\section{Design Implication}
We plan to design a computer mediated tool to create awareness among people about the negative effect of imbalance SNS usage on families. An application can be developed that will keep track of time an individual is spending at different social networks. Specifically, the application should only measure active time spent by an individual on a SNS. If the device remains idle for a while (when no mouse movement or keyboard activity is detected), the application will stop counting the time. An individual may use SNSs from multiple devices such as smart phone or laptop. Thus, to compute the time spent in SNSs by an individual for a certain period of time, the monitoring data needs to be uploaded to a central server~\cite{Reynol}, where only the individual has access to the monitoring data.

Initially, the application will take the information about daily life activities of the individual and her family members, who live at the same places. Based on this information and the individual's SNS usage pattern, the application will be able to give warning message if the individual is using SNSs in an imbalanced way with her family members. The application can analyze the individual's SNS usage pattern for a time period and warns the individual about the level of her imbalance SNS usage per week or month.

\section{Conclusion}
In this paper, we study the impact of imbalance SNS usage on family lives. Our extensive analysis show that imbalance SNS usage has certainly created a family tension. It is true that virtual interactions with family members, especially who lives far away, help to feel them more connected. However, our study reveals that if an individual uses SNSs when one of her family member is expecting to spend time with her, the family member may become annoyed, sad and lonely. This paper is the first attempt to identify the cause and negative effect of imbalance SNS usage. In the future, we aim to conduct a periodic survey to examine how emotions of people change over time with the change of the level of imbalance SNS usage. In addition, we plan to develop a computer mediated awareness tool that can warn if an individual uses SNSs in an imbalanced way with her family member.

\bibliographystyle{acm}
\bibliography{ImpSNS}

\appendix
\gdef\thesection{Appendix \Alph{section}}
\section{}
\label{sample_interview}
In this section, we present 5 interviews. Let the IDs of five anonymous interview participants are A, B, C, D,and E. The demographic information of five participants in the form of $<$age, occupation$>$ are A:$<$52, service holder$>$, B:$<$49, service holder$>$, C:$<$45, service holder$>$, D:$<$29, housewife$>$, and E:$<$52, housewife$>$. In response to the duration of imbalance SNS usage, our participants reacted as follows:
\textbf{Participant A:} ``\emph{It has been a long time that my child is always busy with SNSs when she is at home. I don't think that there is any fixed amount of time.}"\\
\textbf{Participant B:} ``\emph{Whenever my child is at home, he is busy with SNSs, so I guess imbalance is always present.}"\\
\textbf{Participant C:} ``\emph{Yes this type of imbalance is common, but it is not that prominent in my house.}"\\
\textbf{Participant D: }``\emph{Yes, imbalance happens. When he comes to my house, he checks update on his phone. }"\\
\textbf{Participant E:} ``\emph{I don't know how much imbalance occurs, she uses SNSs whenever I am free.}"\\

Thus, the imbalance SNS usage is hard to quantify in terms of time. The participants shared their negative feelings due to the imbalance SNS usage of their family members as follows:
\textbf{Participant A:} ``\emph{I felt annoyed first, sometimes sad thinking that it is creating a distance between us.}"\\
\textbf{Participant B:} ``\emph{I felt annoyed but at times became worried. We have very little communication when he is at home. Yes he sometimes spends time with me though it is a rare occasion.}"\\
\textbf{Participant C:} ``\emph{But when it happens sometimes, I feel annoyed about it. I do not feel sad or lonely since I remain busy in my work.}"\\
\textbf{Participant D:} ``\emph{Maybe he feels bored. But I want him to spend time with my family, talk to them.}"\\
\textbf{Participant E:} ``\emph{I feel lonely sometimes, I go to her room, and find her using SNSs. Sometimes I feel upset because she is not showing me the pictures our relatives share in SNSs. I can't see them for not having a SNS account. Earlier she used to talk to me more and share her daily incidents with me but this is not the same now. }"\\

The interviewer asked the participant about their SNS usage and possible solutions to address the problem of imbalance SNS usage and the participants replied:
\textbf{Participant A:} ``\emph{No I do not have access to any SNS. I tried once but I did not like the idea of SNS. It seemed to me that there is no privacy. I do not think using SNSs in a balanced way will reduce this imbalance for families but might work in some cases.}"\\
\textbf{Participant B:} ``\emph{I also have a SNS account. If balanced usage means both members are using SNSs at the same time, so they do not feel alone then I don't think it's a very good solution. To me, face to face communication is always interesting rather than this virtual communication, specifically with the person I am living with.}"\\
\textbf{Participant C:} ``\emph{When this imbalance occurs, I tell him to leave the device and spend time with other family members. He stops for the time being, and tries to keep my request and spends less time in SNSs. It is like a cycle, after some days I have to warn him again.}"\\
\textbf{Participant D:} ``\emph{Maybe an application where he can use SNSs and still be connected with me would be helpful. A way to send virtual gifts when in SNSs, alert him when using too much, and notify me to comment on his post and vice versa.}"\\
\textbf{Participant E:} ``\emph{No I do not, sometimes I go to her and tell her to spend time with me instead of using SNSs. Then she stops for a while, but I do not think she realizes that herself.}"\\

\section{}

Table \ref{tab:alpha_annoyance},~\ref{tab:alpha_sadness} and~\ref{tab:alpha_loneliness} show the statements that were used to determine the level of annoyance, sadness and loneliness during the survey, respectively. Along with the Cronbach's Alpha, these tables show the percentage of positive responses (agree or strongly agree) for all the individual statements for each emotion.

\begin{small}
\begin{table}[htbp]
\centering
\caption{Questionnaire for Annoyance }
\begin{tabular}{|p {10cm}|c|} \hline
I feel annoyed because & ($\alpha$=0.7521)\\ \hline\hline
My survey partner uses social networking sites when
I am expecting to spend time with my survey partner. & 49.18\% \\ \hline
My survey partner uses social networking sites
for a long period of time continuously. & 59.01\%\\ \hline
My survey partner uses social networking sites
in parallel when s/he is spending time with me, e.g.
gossiping with me and using social networking sites at the same time. & 49.18\% \\ \hline
\end{tabular}
\label{tab:alpha_annoyance}
\end{table}
\vspace{-5mm}
\end{small}

\begin{small}
\begin{table}[htbp]
\centering
\caption{Questionnaire for Sadness }
\begin{tabular}{|p {10cm}|c|} \hline

I feel sad because & ($\alpha$=0.7008) \\ \hline \hline
My survey partner uses social networking sites when I am expecting to spend time with my survey partner. & 27.87\%. \\ \hline
My survey partner used to spend more family time (e.g., watching TV together) before but now uses social networking sites instead. & 55.74\%. \\ \hline
Using social networking sites, my survey partner shows stuffs (e.g. photos) to other family members and can see shared stuffs (e.g. photos) from other family members. I cannot see those stuffs (e.g., photos). & 22.95\% \\  \hline
My survey partner logs on to social networking sites instead of spending time with me when s/he has nothing to do. & 26.23\% \\ \hline
\end{tabular}
\label{tab:alpha_sadness}
\end{table}
\vspace{-5mm}
\end{small}

\begin{small}
\begin{table}[t]
\centering
\caption{Questionnaire for Loneliness}
\begin{tabular}{|p {10cm}|c|} \hline
I feel lonely because & ($\alpha$=0.7725)\\ \hline \hline
My survey partner uses social networking sites when I am expecting to spend time with my survey partner. & 32.79\%. \\ \hline
Using social networking sites, my survey partner shows stuffs (e.g. photos) to other family members and can see shared stuffs (e.g. photos) from other family members. I cannot see those stuffs (e.g., photos). & 11.48\% \\  \hline
My survey partner uses social networking sites for a long period of time continuously. & 18.03\% \\ \hline
My survey partner logs on to social networking sites instead of spending time with me when s/he has nothing to do. & 22.95\% \\ \hline
My survey partner uses social networking sites in parallel when s/he is spending time with me, e.g. gossiping with me and using social networking sites at the same time. & 31.15\% \\ \hline

\end{tabular}
\label{tab:alpha_loneliness}
\end{table}
\end{small}
\vspace{-5mm}


\end{document}